\documentstyle[aas2pp4]{article}

\lefthead{Oppenheimer et al.}  \righthead{Spectrum of Gliese 229B}

\begin{document}

\title{The Spectrum of the Brown Dwarf Gliese 229B}

\author{B. R. Oppenheimer, S. R. Kulkarni, K. Matthews} 
\affil{Palomar Observatory\\ 105-24, California Institute of
Technology\\ Pasadena, California 91125}
\authoremail{bro@astro.caltech.edu}
\bigskip
\centerline{ \&\ }

\author{M. H. van Kerkwijk} 
\affil{Institute of Astronomy\\ University of Cambridge\\ Madingly Road\\
Cambridge, CB3 0HA, UK}

\begin{abstract}

We present a spectrum of the cool ($T_{\rm eff} = 900$ K) brown dwarf
Gliese 229B.  This spectrum, with a relatively high signal-to-noise
ratio per spectral resolution element ($\gtrsim 30$), spans the
wavelength range from 0.837 $\mu$m to 5.0 $\mu$m.  We identify a total
of four different major methane absorption features, including the
fundamental band at 3.3 $\mu$m, at least four steam bands, and two
neutral cesium features.  We confirm the recent detection of carbon
monoxide (CO) in excess of what is predicted by thermochemical
equilibrium calculations.  Carbon is primarily involved in a chemical
balance between methane and CO at the temperatures and pressures
present in the outer parts of a brown dwarf.  At lower temperatures,
the balance favors methane, while in the deeper, hotter regions, the
reaction reverses to convert methane into CO.  The presence of CO in
the observable part of the atmosphere is therefore a sensitive
indicator of vertical flows.  The high signal-to-noise ratio in the 1
$\mu$m to 2.5 $\mu$m region permits us to place constraints on the
quantity of dust in the atmosphere of the brown dwarf.  We are unable
to reconcile the observed spectrum with synthetic spectra that include
the presences of dust.  The presence of CO but lack of dust may be a
clue to the location of the boundaries of the outer convective region
of the atmosphere: The lack of dust may mean that it is not being
conveyed into the photosphere by convection, or that it exists in
patchy clouds.  If the dust is not in clouds, but rather sits below
the outer convective region, we estimate that the boundary between
outer convective and inner radiative layers is between 1250 K and 1600
K, in agreement with recent models.

\end{abstract}

\section{Introduction}

Gliese 229B is the first sub-stellar object outside the solar system
with an effective temperature, T$_{\rm eff}$, well below 1800 K, the
minimum T$_{\rm eff}$ for stars, found using direct imaging techniques
(Nakajima et al. \cite{nak95}).  Other sub-stellar companions have
been inferred to exist through indirect techniques (Wolszczan \&\
Frail \cite{wf92}; Mayor \&\ Queloz \cite{mq95}; Butler et
al. \cite{but97}).  The astrophysical nature of all these objects can
only be understood in detail through spectroscopic studies.
Unfortunately current technology precludes the spectroscopic study of
any of these objects, except Gliese 229B.  Interestingly, the spectrum
of a gaseous, sub-stellar object is primarily determined by its
effective temperature, and secondarily by its surface gravity and
composition (Burrows et al. \cite{bur97}; Burrows et
al. \cite{bur96}).  Thus, the spectrum of Gliese 229B is expected to
resemble the spectra of some of the less massive, but similarly hot,
giant planets found in the radial velocity studies.

Here we present the spectrum of Gliese 229B with high signal to noise
ratio from 0.837 $\mu$m out to 5 $\mu$m.  Gliese 229B emits
approximately 65\% of its emergent flux in this wavelength region.  An
effort like this can perhaps be seen as a precursor to extrasolar
planet studies or an extension of the work of hundreds of planetary
scientists who have studied and modelled the atmospheres of the
planets orbitting the Sun.  In addition, these observations also give
extremely interesting insights into the atmospheric physics of brown
dwarfs, including the atmospheric dynamics.  Our now detailed set of
observations can be compared with recent models.

\section{Observations and Data Reduction}

All observations reported here were made at the Keck Telescopes.  At
these telescopes, a major problem in observing Gliese 229B is the
presence of diffraction spikes from Gliese 229A.  The image of a star
produced by one of the Keck telescopes has six diffraction spikes
spaced every 60\arcdeg\ around the center of the image.  These spikes
are produced by both the secondary support vanes and the primary
mirror's segment gaps.  The spikes are fixed with respect to the
horizon.  In celestial coordinates, the position angle of any spike is
the paralactic angle plus an integer multiple of 60\arcdeg\ (see
right-top panel of Figure 1).  As a result, the spikes
rotate as the star transits the sky.  Useful spectra of Gliese 229B
could only be obtained when the diffraction spikes of the very much
brighter Gliese 229A were $\ga\!15^\circ$ away from the position angle
of Gliese 229B.  This reduced the time during which it was possible to
observe Gliese 229B to 1.5 hours just after transit.  Since Gliese
229B is not visible on the slit-viewing camera, we set up on Gliese
229A and applied a blind offset from there (7\farcs 78 at position
angle 161\fdg 3; Nakajima et al. \cite{nak95}).

\subsection{Optical Data}\label{sec:lris}

We observed Gliese 229B using the Keck I telescope on 21 January 1996
(UT) with the Low Resolution Imaging Spectrometer (LRIS; Oke et al.
\cite{oke95}).  We made three exposures of 1800 s duration each, using
a Tektronix 2048$\times$2048 pixel CCD with 24 $\mu$m pixels.  We used
the 1200 line mm$^{-1}$ grating, blazed at 750 nm to cover the
wavelength region of 847 nm to 973 nm at 0.06 nm per pixel.  A
1\arcsec\ slit, corresponding to 4.65 pixels, gave a spectral
resolution of about 0.3 nm.  The slit was placed East-West for all of
the LRIS observations.  An OG570 filter suppressed the second order
spectrum.

On 6 February 1997 (UT), three more 900 s LRIS exposures were taken in
the middle of the interval when the diffraction spikes were more than
$15^\circ$ away.  The set-up was the same, except that the instrument
had been moved to Keck~II, and we used a new, more efficient 831 line
mm$^{-1}$ grating, blazed for a wavelength of 810 nm, covering the
wavelength range 837 nm to 1024 nm at 0.09 nm per pixel.  With the
1\arcsec\ slit, the resolution was about 0.4 nm.  On both nights the
seeing was generally sub-arcsecond ($\sim$ 0\farcs 8) and a few thin
cirrus clouds sailed about in the sky.  The spectra from these two
observations were combined by smoothing the higher resolution spectrum
to the 0.4 nm resolution and adding them.  Full details on the
reduction technique follow.

The spectral data were reduced using MIDAS and programs running in the
MIDAS environment.  The presence of Gliese 229A's diffraction spikes
in the spectral images required special care during the reduction.
The approach we used is illustrated in Figure 1.  The
top panel shows a typical spectral image, after bias-subtraction and
flat-field correction.  First we removed the telluric emission lines
(mostly OH).  For this purpose, a third-degree polynomial was fitted
through every column in regions free of stellar light (indicated by
`S' on the right-hand side of the panel).
 
The subtraction of the strongest sky lines is not always satisfactory
(see second panel of Figure 1).  Most likely, this
results from the fact that interference fringing in the CCD is not
corrected adequately for the sky emission lines, as the response of a
given pixel to monochromatic light is different from the response to
the continuum emission from the flat-field lamp.  The resulting
structure over a column has a relatively large scale, however, and is
mostly removed in the next step, in which the scattered and diffracted
light from Gliese 229A is subtracted by fitting a second-degree
polynomial to two small regions adjacent to the spectrum of Gliese
229B (see Figure 1).  As the spectra are not aligned
perfectly, these regions are chosen slightly differently for different
parts of the chip.
 
One-dimensional spectra were extracted using a scheme similar to the
``optimal extraction'' technique of Horn (\cite{horn:86}), in which an
average stellar profile along the slit is created to estimate the
total flux from each pixel in a given column.  All estimates are
combined using weights inversely proportional to the square of each
pixel's formal uncertainty (determined from the read noise and Poisson
noise in the raw data). In the lower panel of Figure 1, we show the
difference between the raw data and the extracted data (in standard
deviations).  It is apparent that the extraction is quite accurate,
since very little structure remains.  We also extracted the spectrum
of Gliese 229A with the same technique, after removing the
contribution from Gliese 229B from the image.  This allows us to
verify that no contamination from Gliese 229A remains in Gliese 229B's
spectrum.  Indeed Figure 6 shows that none of the
features in A are present in the spectrum of B.
 
Wavelength calibration employed argon and krypton/mercury arc spectra
taken after the series of exposures.  For the 1200 line mm$^{-1}$
spectra, only 7 lines were available, but they were quite evenly
spread over the wavelength range.  A fit to these lines's positions
with a third-degree polynomial left a root-mean-square residual of
0.023 pixel.  For the 831 line mm$^{-1}$ spectra, 11 lines were used
(achieving an rms of 0.03 pixel), but none of these were longward of
970 nm.  However, a comparison of the locations of OH airglow lines in
our spectral images with the wavelengths given by Osterbrock et
al. (\cite{ostfb:97}) shows that the wavelength calibation is accurate
to better than 0.1 nm everywhere on the spectrum.  For removal of
telluric water vapour lines as well as relative flux calibration, we
used spectra of Feige~34 (sdO; $V=11.2$; Massey \& Gronwall
\cite{massg:90}), taken after Gliese 229B at similar airmass.  The
final absolute flux calibration was achieved using the photometric
results in Matthews et al. (\cite{matt96}).

\subsection{Infrared Data}\label{sec:nirc}

On the nights of February 8 and 9, 1997 (UT) we obtained the Z, J, H,
K, L and M band spectra using the Near Infrared Camera (NIRC; Matthews
\&\ Soifer \cite{mat94}) at the Cassegrain focus of the Keck I
telescope.  These spectra were obtained in several stages using the
grism mode of the camera.  The camera permits recording of
low-resolution spectra in the Z, J and H bands simultaneously, then
the H and K bands and finally the L and M bands.  The ``ZJH'' region
was obtained in 10 exposures of 40 s of ``on-source'' exposure time.
After every 40 s exposure, we shifted the telescope 15\farcs 6 North
and took a 40 s long ``sky'' frame.  The slit was placed East-West, as
it was during the LRIS observations.  This produced pairs of
``on-source'' exposures taken right before corresponding ``sky''
exposures.  The distance 15\farcs 6 was chosen so that we would be
sampling exactly the same sky background relative to the primary star
(Gliese 229A), but on the opposite side of the star.  This seemed to
be the optimal way to remove the point-symetric background.

We used the grism with 150 line mm$^{-1}$ and a slit 4.5 pixels wide
(0\farcs 675) to obtain the ``ZJH'' spectrum.  Blocking filters were
in place so that only light from the Z, J and H bands reached the
detector.  The H band spectrum was obtained a second time in a series
of 22 more 40 s exposures (plus 22 corresponding sky exposures) in
which the H and K spectra were recorded.  In these spectra, the 120 line
mm$^{-1}$ grism was in place with corresponding blocking filters as
well.  The weather was clear and photometric and the seeing was
0\farcs 7 on both nights.  

The L and M band spectra were slightly trickier to obtain.  As far as
we know this work represents the first time the Keck telescope has
been used for 3 $\mu$m to 5 $\mu$m spectroscopic observations.  As a
result we had to ascertain which electronics settings were optimal and
what readout times and integrations were necessary.  
As shown by the results presented here, NIRC is fully capable and rather
sensitive in these wavelength bands.

All of the spectral images contained both Gliese 229A and B, due to
the diffraction spikes of A.  Data reduction involved the following
steps.  For each ``on-source'' frame (which entailed the coaddition by
the NIRC electronics of 1000 individual array readouts)
we subtracted the ``sky'' frame immediately following.  We then
extracted spectra from bars ten pixels in width across the slit of
both Gliese 229B and A.  Gliese 229B's spectral signature was quite
detectable in the single, ``sky-subtracted'' images.  The extraction
technique involved the ``optimal extraction'' method in which the
pixels are summed in the spatial direction but weighted according to
the signal in each pixel.  The technique is similar to that used for
the optical data, and it leads to a significant improvement in
signal-to-noise ratio over simple addition.  To make sure the sky
background and particularly any contamination from A were not present
in our final spectrum, we subtracted the corresponding ``optimally
extracted'' spectrum ten pixels to either side of the spectral region
of B.

We removed the instrumental response by dividing the extracted
spectrum by the extracted spectrum of a calibrator star.  In the case
of the Z, J, H and K bands this calibrator was SAO 175039, a G8 star
which smoothly approximates a black body spectrum through this
wavelength region, except for a few hydrogen absorption lines which we
edited out by linear interpolation.  In the case of the L and M bands,
we used Gliese 229A, an M1V star, which is quite flat in the region
redward of 2 $\mu$m.  (For L and M spectra of other early M dwarfs,
see Berriman \&\ Reid \cite{reid87} or Tinney et al. \cite{tin93}.)
Ideally one should use a calibrator star such as a bright F or G type
star.  However, the use of A is sufficient in this case.  Absolute
flux calibration involved matching the flux density in the spectrum
with the photometric measurements of Matthews et al. (\cite{matt96}).

The wavelength calibration of the infrared data involved a simple
method.  We used the edges of the filter bands, clearly visible in the
calibrator star spectra, to determine the wavelengths of the edges of
each segment of the spectrum.  The 50\% transmission points of the
bands are known to an accuracy of 10 nm.  With these ``edge
wavelengths'' known we linearly interpolated to determine the
wavelength for each pixel across the spectrum.  This introduces errors
on the scale of 1 nm.

Finally we added all of the extracted, calibrated spectra together.

Figure 2 compares our 1 $\mu$m to 2.5 $\mu$m spectrum with the higher
resolution spectrum of Geballe et al. (\cite{geb96}).  The plot shows
that there is good agreement between the spectra.  The differences,
which are most pronounced in the 1 $\mu$m to 1.1 $\mu$m and the 1.6
$\mu$m to 2 $\mu$m regions, amount to no more than 10\%\ of the
signal.  These discrepancies are attributable to possible minor
contamination from Gliese 229A in the 1.6 $\mu$m to 2.0 $\mu$m region
of our spectrum as well as some minor differences in wavelength
calibration.  Alternatively, the Geballe et al. (\cite{geb96})
spectrum might have been affected by atmospheric dispersion, causing
some of the bluer wavelegths to be slightly reduced in brightness
since some of the blue light would have been blocked by the slit.  In
Geballe et al. (\cite{geb96}) it is shown that the small features in
the brighter parts of the spectrum are almost entirely real and
correspond closely to known absorption lines in water.  Our spectrum
is unable to resolve these features.  However, the agreement between
the spectra is an excellent confirmation of the data, as well as an
indicator that our reduction method works and has no major
contamination from scattered light from Gliese 229A.

Figure 3 shows the relative instrumental response of
NIRC with the 60 line mm$^{-1}$ grism in the L band plotted along with
our spectrum of Gliese 229B.  The absorption band at about 3.4 $\mu$m
is due to the resin inside NIRC.  This resin is applied to the surface
that will become the grating.  Then a master grating is pressed
against the resin so that an imprint of the grating is left in the
resin, which subsequently dries and becomes the ``replicant'' grating
used in NIRC.  The instrumental transmission curve is simply the
extracted spectrum of Gliese 229A, which, as we mentioned above, is
flat in this region. The final spectrum of B shows none of the
features due to this instrumental response.

\section{The Spectrum}

The complete spectrum is presented in Figure 4, with the
most important features labelled.  The 1 $\mu$m to 2.5 $\mu$m part of
this spectrum has been discussed in detail in Oppenheimer et
al. (\cite{opp95}) and Geballe et al. (\cite{geb96}).  Hence we
concentrate on the new short and long wavelength parts.  In the latter
we find the fundamental methane band at 3.3 $\mu$m, and an interesting
small feature in the M band which we attribute to CO, in confirmation
of the claims of Noll et al. (\cite{nol97}; see section \ref{sec:co}).
In the optical part, shown enlarged in Figure 5, there
are two prominent atomic lines due to cesium, a very weak feature
caused by methane and a strong steam band.  These are all discussed in
detail in the following sections.  \notetoeditor{This figure
{fig:spect} should be printed large in the journal.  It is the central
result of the paper and the details in the spectrum are important.
Perhaps using a full page would be appropriate.}

Before discussing the detailed parts of the spectrum it is useful to
compare the spectrum of Gliese 229B with spectra of some of the lowest
mass stars.  Figure 6 shows the optical part of the
spectrum compared with that of vB 10 and LP 944$-$20, which were
studied by Kirkpatrick et al. (\cite{kp97}).  Note the distinct
absence in Gliese 229B of any of the molecular absoprtion features due
to refractory elements, such as VO and TiO.  Also the steam feature is
far deeper and more exaggerated in Gliese 229B.  In contrast to the
stars, Gliese 229B seems smooth blueward of the steam feature, with
the exception of the two neutral cesium lines and a possible faint
methane feature.  (See \S \ref{sec:ces} and \S \ref{sec:meth} for a
detailed discussion.)

Figure 7 compares Gliese 229B's near IR spectrum with
those of GD 165B and vB 10 again.  The comparison spectra are
reproduced here courtesy of H. R. A. Jones and appeared in Jones et
al. (\cite{jon94}).  The huge swings in flux density in Gliese 229B
are not to be found in either star.  In particular, this figure shows
the great importance that methane has on Gliese 229B's spectrum,
causing drastic drops in the flux density at 1.6 $\mu$m and 2.2
$\mu$m.

The water bands at 1.1 $\mu$m and 1.4 $\mu$m are also vastly deeper in
Gliese 229B than in the stars.  More specifically, the water band at
1.4 $\mu$m is about 2.5 times deeper in Gliese 229B than it is in GD
165B.  Some of these differences may be magnified by the presence of
dust in GD 165B and vB 10.  Dust serves to smooth out spectra such as
these, and it has recently come to light that dust is quite important
in the spectra of some very low mass stars (Jones \&\ Tsuji
\cite{jon97}).  (See \S \ref{sec:dust}.)

Interesting comparisons can be made of the Gliese 229B spectrum with
that of Jupiter.  Although the spectrum of Jupiter is due to reflected
sunlight throughout the wavelengths discussed here, there exist many
similarities.  We discuss these in the sections below where
appropriate.

What follows is a detailed explanation of various aspects of Gliese
229B which can be addressed with the data presented in this paper.

\section{Methane}\label{sec:meth}

Our spectrum in the 3 $\mu$m to 4 $\mu$m region shows an extremely
large absorption feature (Figure 4).  We attribute this
to the fundamental methane absorption band, which has been observed in
planets and moons in the solar system.  Noll (\cite{nol93}) detected
an interesting spike in the spectrum of Jupiter at 3.52 $\mu$m.  Other
researchers have also noticed this feature (Drossart et al.
\cite{dro96}).  We do not see this feature in our spectrum of Gliese
229B.  Drossart et al. (\cite{dro96}) attribute this feature to
reflected solar light emerging between what is actually two important
methane features.  The principal reason we cannot comment on the
presence of this feature in Gliese 229B is that the signal to noise
ratio in this part of the spectrum is exceedingly low.  For that
reason the depth of this band cannot be determined from this data.

Another formerly undetected band of methane presented here is the
faint band in the 0.89 $\mu$m region.  This band, which is quite
prominent in the reflectance spectra of Jupiter, Saturn, Titan, Uranus
and Neptune, has been shown by Karkoschka (\cite{kar94}) to be
strongly dependent upon the effective temperature.  Large differences
in the depth of the band are apparent in the spectra he presents.
This is important because the temperature dependence of the absorption
coefficient for methane is poorly known.  Examination of the results
in Karkoschka (\cite{kar94}) show that this band should be extremely
weak at high temperatures, such as the 900 K temperature of Gliese
229B.  Thus, the weakness of this feature is expected.

\section{Water}

Water is by far the most important source of non-continuum opacity in
Gliese 229B.  (Continuum opacity comes, as indicated by models, from
H$^-$, H$_2^-$ and H$_2$ collision induced absorption; Burrows et
al. \cite{bur97}.)  In models of Gliese 229B, the absorption between
the near IR filter bands (which are naturally defined by telluric
water absorption) are broad and deep.  This is highly beneficial.  It
means that the majority of the flux from Gliese 229B and any similar
objects, such as extrasolar giant planets, will pass through the
earth's atmosphere (Matthews et al. \cite{matt96}), and that brown
dwarf or extrasolar planet studies are not hindered by this telluric
absorption.

Geballe et al. (\cite{geb96}) showed that the narrow features in the
2.0 $\mu$m to 2.2 $\mu$m region correspond closely to the opacity of
water.  Saumon et al. (1996) have shown that a measurement of the
surface gravity of Gliese 229B could be made using these very fine
water features.

The optical spectrum (Figure 5) shows many narrow
features redward of 0.925 $\mu$m.  These are essentially all real
features and are a result of the complex structure of the water
molecule (F. Allard, private communication).  In principal one can
conduct an analysis of these water features similar to that in Meadows
\&\ Crisp (\cite{mead96}), in which the thermal structure of the
atmosphere of Venus was deduced.  However, this entails modelling of
line profiles and spectral synthesis with detailed radiative transfer
models, which our group is unable to do at present.  Unfortunately the
models for Venus cannot be used because Gliese 229B (presumably) has
no rocky surface underlying its atmosphere.  In the case of Venus, the
surface has a profound impact on the radiative processes in the
atmosphere (Meadows \&\ Crisp \cite{mead96}).

\section{Carbon Monoxide}\label{sec:co}

The first vibration-rotation band (1-0) of CO is centered near 4.7
$\mu$m.  Motivated by a paper by Fegley \&\ Lodders (\cite{nol96}), which
discussed the importance of the detection of CO, we measured the 3
$\mu$m to 5 $\mu$m spectrum of Gliese 229B presented here.  Since
then, the presence of CO in the atmosphere of Gliese 229B has been
reported by Noll et al. (\cite{nol97}; hereafter NGM) NGM, using
spectral data in the 4.7 $\mu$m region, claim to detect the expected
increase in flux density in the center of the 1-0 vibration-rotation
band.  Though the noise in their data was appreciable, the detection
is clear and the model spectrum they generated agrees well with the
data.

Our data confirm their result, as can be seen in Figure 8,
where we have reproduced NGM's Figure 2 (courtesy of Mark Marley),
showing their data and model, but with our data overdrawn.  Their
model fits the observations best with a mole fraction, q, of qCO = 200
ppm (parts per million) and a brown dwarf radius of $8\times 10^4$ km
(which affects the flux density as well as the model's adiabat).  We
have scaled the flux density of the NGM data by a factor of two to
match the flux density we measured in Matthews et al. (\cite{matt96}).
The uncertainty on each of the points in our data is approximately
\twothirds\ the uncertainty in the data of NGM.  Our data fit the
model extremely well, even in regions where the NGM data do not.

NGM state that the sudden break at 4.82 $\mu$m in their spectrum is
due to incorrect removal of a telluric line.  Indeed, we do not see
this feature in our spectrum.

As NGM discuss, one does not expect CO to be present in the atmosphere
of Gliese 229B in thermochemical equilibrium.  Fegley \&\ Lodders
(\cite{feg96}) computed models of the thermochemical equilibrium
abundances of many molecules for a pressure temperature profile of
Gliese 229B from Marley et al. (\cite{mar96}).  They found that only at
a temperature of about 1400 K would one expect the mole fraction of CO
to equal that of CH$_4$.  At 1250 K CO is about 10 times less abundant.
At the 900 K isotherm one would expect the abundance of CO to be almost
10$^{-4}$ times that of CH$_4$.  In the atmospheres of Jupiter and
Saturn, the qCO/qCH$_4$ = 1 boundary is much deeper (Fegley \&\
Lodders \cite{feg94}).  The only explanation for the presence of CO in
the photosphere of Gliese 229B is that it is not in thermochemical
equilibrium.  This seems to require convection which would bring
measurable quantities of CO up from the deeper and warmer parts of the
atmosphere.  The convection need not necessarily reach all the way to
1400 K.  Even if it extended into the 1250 K region, substantial
quantities of CO could be pulled up into the outer atmosphere
(Fegley \&\ Lodders \cite{feg94}, Fegley \&\ Lodders \cite{feg96}).

This indicates that the outer convective region of Gliese 229B is at
least as deep as the 1250 K isothermal surface.  Recent models by
Burrows et al. (\cite{bur97}) found that an object like Gliese 229B
has an outer radiative region, followed as one progesses inward by a
thin convective region which spans a temperature range depending upon
the mass and effective temperature.  Below this thin outer convective
region lies another radiative region which extends in almost all of
the models to a depth where the temperature is about 2000 K.  Inside
this inner radiative region is a final convective region which reaches
to the core of the brown dwarf.  This concept of shells of alternating
convective and radiative regions seems to be in agreement with the
observations made here.  The CO might form easily in the outer
convective shell, which can reach to temperatures of over 1250 K, and
get dredged up to the observable part of the atmosphere.  To make this
more clear, one must consider the presence of dust in the spectrum of
Gliese 229B.

\section{Dust}\label{sec:dust}

The issue of dust in Gliese 229B has been the subject of some
discussion, particularly among the modellers of brown dwarf and planet
atmospheres (e. g. Marley et al. \cite{mar96}, Tsuji et
al. \cite{tsu96}, Allard et al. \cite{all96}).  At some point in the
near future, the application of a detailed theory of the cloud
microphysics (such as appears in Rossow \cite{ross}) may be possible.
For now, we restrict ourselves to a somewhat more qualitative
discussion.

Metallicity of Gliese 229B is important to any discussion of its dust
content.  At this time we have no direct measurment of the metallicity
of Gliese 229B.  However, the two measurements of the metallicity of
Gliese 229A are [M / H] = +0.15 $\pm$ 0.15 (Mould 1978) and [Fe / H] =
-0.2 $\pm$ 0.4 (Schiavon et al. 1997), which are both consistent with
solar metallicity (within the errorbars).  If Gliese 229B formed the
way planets are thought to form, out of a circumstellar disk
surrounding the nascent Gliese 229A, it may be that the metallicity of
the brown dwarf is higher than that of its sun, as is the case for the
outer planets of the solar system.  In this case, one would expect to
see even more dust (Burrows et al. \cite{bur97}).  On the other hand,
if Gliese formed out of the collapse of an interstellar cloud, the way
Gliese 229A presumably formed, the metallicities should be identical.
This is important because of the apparent lack of dust influences in
the spectrum, as we discuss below.

Jones \&\ Tsuji (\cite{jon97}) have demonstrated that dust plays an
extremely important role in the spectra of the very lowest mass stars.
Looking at progressively cooler stars Jones \&\ Tsuji (\cite{jon97})
find that dust increases in importance past the spectral
classification M6.  The evidence for this is that the prominent TiO
and VO bands that define late M type spectra suddenly seem to reduce
in strength past the M6.5 V spectral type.  This cannot be reconciled
with dust-free models of late type stars, which show increasing
strengths of the TiO and VO bands until the bottom of the main
sequence (Allard \&\ Hauschildt \cite{all95}).  Furthermore, Tsuji et
al. (\cite{tsu96a}) have shown that GD 165B's spectrum can only be
explained with the presence of large quantities of dust.  It may be
that past the M6.5 V spectral type stars do not all appear the same:
some may be rather dusty, while others may not.  This might be due to
differences in metallicity and in details of the stars's weather
patterns.  Observational support for such differences comes from the
diversity of spectra observed by Kirkpatrick et al. (\cite{kp97}).

The results of Jones \&\ Tsuji (\cite{jon97}) would lead one to expect
that a substellar object such as Gliese 229B would be even more
affected by dust.  However, based on the photometry of Gliese 229B (as
reported by Matthews et al. \cite{matt96}), Tsuji et
al. (\cite{tsu96}) concluded that dust does not play a role in
determining the spectral energy distribution.  From our high
signal-to-noise ratio spectral data through the deepest parts of the
methane absorption bands, we can more quantitatively compare the
observed spectrum with a range of spectra calculated with differing
quantities of dust.  We find that the spectrum presented here cannot
be reconciled with any of the dusty models that Tsuji et
al. (\cite{tsu96}) computed.  Even the least dusty model of Tsuji et
al. (\cite{tsu96}) does not fit the observed spectrum.  For instance,
the observed factor of 10 drop in flux density in the K band matches
only the dust free models of Tsuji et al. (\cite{tsu96}).  The dusty
models of Tsuji et al. (\cite{tsu96}) show a pronounced dampening of
this methane absorption band.

The question that now arises is ``Where is the dust that is expected
to be present in the brown dwarf?''  We see two possibilities.  The
first, suggested by Tsuji et al. (\cite{tsu96}), is that the lack of
dust is only apparent; the dust, which one expects from thermochemical
considerations, is present but in clouds in the photosphere with a
small covering fraction.  This would presumably be an observably
testable suggestion.  One could look for variability in the absorption
bands of Gliese 229B.  In the various spectral images we have taken at
Keck and Palomar we found no evidence for variability greater than
10\%\ on time scales from several minutes to two days and over a one
year baseline in the 1.6 $\mu$m methane absorption band, the depths of
which are easily detected in a single 40 second exposure on Keck.
Whether the presence of clouds would produce a 10 \%\ or larger effect
is not clear.  However, this is probably a viable explanation.

Another possible reason for the lack of dust is that it sits deep
within the atmosphere below the outer convective region in an inner
radiative region (first predicted in planets by Guillot et
al. \cite{gui94}, and predicted to exist in brown dwarfs by Burrows et
al. \cite{bur97}), so that it is not blown back up into the
photosphere through convection.  There are several points we can raise
in support of this idea.

We argued above (\S \ref{sec:co}) in confirmation of Noll et
al. (\cite{nol97}) that there is an appreciable quantity of CO visible
in the spectrum of Gliese 229B, and that this is many orders of
magnitude higher than the amount expected from thermochemical
equilibrium.  We further argued that this can be the result of
convection in the brown dwarf's atmosphere.  This seems to imply that
the convective region of the brown dwarf must extend at least to a
depth where CO is abundant.  Fegley \&\ Lodders (\cite{feg96}) showed
that the CO to CH$_4$ phase transition ought to happen at around 1400
K and a pressure of about 10 bars in Gliese 229B.  As mentioned above,
it is possible that the CO we observe comes from a lower temperature
where it exists but in smaller relative quantities.  Thus it could
perhaps come from as low a temperature as 1250 K, where the mole
fractions of CO and CH$_{4}$ differ by a factor of 10.  We already
know, though, that the dust forms at much higher temperatures than
this.  By far the most abundant refractory elements are Fe, Mg, Si and
Al.  These form Al$_2$O$_3$ (corundum) in the late M dwarf regime, Fe
clusters at 2000 K (Fegley \&\ Lodders \cite{feg96}), and MgSiO$_3$
(enstatite) at 1600 K (Tsuji et al. \cite{tsu96a}, Sharp \&\ Huebner
\cite{sha90}).

This may indicate where the boundary between the outer convective and
inner radiative regions of the brown dwarf are.  Specifically, the
boundary must be at a temperature above 1250 K, to permit dredging of
substantial quantities of CO into the cooler regions, and yet cooler
than 1600 K, where enstatite forms.  If the outer convective region
extends deeper and hotter than 1600 K then the condensation and
falling or raining rate of enstatite must be much faster than the
convective wind speed, so that the enstatite and other types of dust
never reach the outer atmosphere.  This process would be similar to
the falling of rain droplets in the Earth's convective and stormy
atmosphere.  Rain falls despite the convection.  Presumably one could
model this process to answer the question raised in this paragraph.
If the dust is below the outer convective region in a radiative
region, clouds are not needed to reconcile the observations with
theory.

According to Marley et al. (\cite{mar96}) and Marley (\cite{mar97}),
theoretical temperature versus pressure curves for Gliese 229B show
that the atmospheric structure becomes adiabatic at approximately 1600
K and 10 bars of pressure.  This is consistent with our suggestion
above, that the outer convective/inner radiative boundary is between
1250 K and 1600 K.  It is also consistent with the new models by
Burrows et al. (\cite{bur97}).

If this is true, then one would predict that objects at some
temperature below the 1800 K of GD 165B will begin to show the effects
of dust less and less.  Indeed, the expectation is that the much
hotter brown dwarfs will exhibit dust in their spectrum.  The possibly
close proximity of the radiative-convective boundary in Gliese 229B to
the formation temperature for enstatite suggests that a brown dwarf
even slightly hotter than Gliese 229B might show some effects of dust
in its spectrum.  Thus, one should see a progressive decrease in the
importance of dust through the range of effective temperatures from
1800 K to 900 K.  Alternatively, below a certain temperature, the
photosphere may become dust free.  This could happen if convection is
unable to bring the dust back into the photosphere because it just
rains out of the convective region.

It is entirely possible, however, that the type of internal structure
we discuss here is largely model dependent.  For example, additional,
as-yet-unidentified opacity sources inside the brown dwarf atmosphere
could upset or even drastically change the conclusions of the models
that there are inner radiative and convective zones.

As a final note on the dust in Gliese 229B, we remark upon the
smooth-looking parts of the optical spectrum.  This may be due to a
haze of exotic types of dust not yet identified with extremely fine
particle size, but this requires additional work on behalf of the
modellers (M. Marley, F. Allard, personal communication).

\section{Cesium}\label{sec:ces}

Of the alkali metals (Li, Na, K, Rb, Cs and Fr), only lithium, sodium,
potassium and rubidium have been known to play an important role in
low mass star spectra (Kirkpatrick et al. \cite{kp97}; Basri \&\ Marcy
\cite{bm95} and references therein).  Recently, Tinney (\cite{tin97})
found cesium absorption lines in spectra of several very low-mass
stars.  This completes the set of alkali metals one would expect to
find in such stars.

The reason the alkali metals are important is that they have very low
ionization potentials: from 5.390 eV for Li to 3.893 eV for Cs
(Letokhov et al. \cite{let87}).  These are the lowest ionization
potentials for any of the elements.  As a result, in most stellar
atmospheres they are ionized and would only be visible in an
ultraviolet spectrum, a wavelength region never used to classify M
dwarfs.  As one considers stellar atmospheres of cooler and cooler
temperatures, the neutral alkalis begin to appear, with Cs appearing
only in the very coolest atmospheres.  Continuing this progression
through cooler atmospheres, as one crosses into the brown dwarf
regime, these metals should begin to form molecules.  Thus, they would
again disappear from the optical to near IR spectra.

In the case of Gliese 229B, the two strongest cesium lines are
present.  (See Figure 5.)  The other alkali metals have
their strongest lines blueward of the shortest wavelength we were able
to measure.  (Li is at 670.8 nm and 812.6 nm; Na is at 589.5 nm and
589.0 nm; K is at 769.9 nm and 766.5 nm and Rb is at 794.8 nm and
780.0 nm.)  Unfortunately we are unable to use the lines for a
reliable calculation of the abundance of cesium in Gliese 229B.  This
is beacuse there exist to date no curve of growth models for these
absorption lines.

The cesium lines present in the optical region of the spectrum of
Gliese 229B comprise the principal doublet of cesium.  The line at
852.1 nm corresponds to a transition of the solitary valence electron
from the 6 $^{2}$S$_{1 \over 2}$ state to the 6 $^{2}$P$_{3 \over 2}$
state.  The other line at 894.3 nm corresponds to the transition from
6 $^{2}$S$_{1 \over 2}$ to 6 $^{2}$P$_{1 \over 2}$.  (This notation
follows the standard spectroscopic notation convention,
$^{2S+1}L_{J}$.)  Kirkpatrick (private communication) has found the
same two cesium lines in a new spectrum of GD 165B, as well.  These
are by far the strongest lines of cesium.  The next strongest line at
917.2 nm is several times weaker than either of these and is not
visible in Gliese 229B.

An important question to answer is ``At what temperatures can we
expect these alkali metals to form molecules?''  This question, which
requires thermochemical equilibrium calculations, has been thoroughly
addressed for planet sized objects like Jupiter by Fegley \&\ Lodders
(\cite{feg94}).  Their thermochemical equilibrium calculations would
lead one to believe that none of the alkali metals should be visible
in the spectrum of Jupiter, because they ought to have formed
molecules already.  Indeed, in Jupiter, one expects CsCl to dominate
over Cs even at temperatures as high as 2000 K.  (The brightness
temperature in the region of the Cs lines in Gliese 229B is
approximately 1300 K.)  However, it is not entirely correct to use the
calculations for Jupiter when discussing Gliese 229B.  The adiabat for
Gliese 229B places a given temperature at a substantially lower
pressure.  A. Burrows (personal communication) believes that the turn
over from Cs to CsCl in Gliese 229B happens around 1500K.  However,
until the full calculations are carried out, reconciling the theory
with this detection of atomic Cs cannot happen.  It is possible that
one merely expects the deep Cs lines we see, but if they are not
compatible with thermochemical equilibrium, it is possible that some
quantity of Cs is being convected into the observable atmosphere much
the way the CO is, as we described in \S \ref{sec:co}.  The
thermochemical quilibrium calculations could also lead to an important
temperature diagnostic for brown dwarfs and very low-mass stars.  By
observing which alkali metals exist in atomic form, one could
determine the temperature of a given object through comparison with
the thermochemical equilibrium calculations.

These considerations also have implications for the so-called lithium
test for brown dwarfs.  It may be that the lithium test, in which one
attempts to detect atomic lithium in the candidate brown dwarf (Rebolo
et al. \cite{reb92}; Magazz\`{u} et al. \cite{mag93}), might only be
meaningful for the highest temperature brown dwarfs.  The temperature
at which lithium becomes LiOH determines this.  For the case of the
Jupiter adiabat, this is well above 2000 K (Fegley \&\ Lodders
\cite{feg94}).  The field would benefit from new thermochemical
equilibrium calculations using the adiabats found for Gliese 229B
(Marley et al. \cite{mar96}; Burrows et al. \cite{bur97}).

\acknowledgments The authors would like to thank Hugh Jones and Davy
Kirkpatrick for supplying their data for the purposes of comparison
with Gliese 229B.  We would also like to thank Davy Kirkpatrick for
lengthy discusions on atomic and molecular identifications, Mark
Marley for helping with Figure 8 and for excellent
discussions of convection and weather, Bruce Fegley for a very
thorough referee's report and Neill Reid for a very useful and
thorough reading of the manuscript.  SRK graciously thanks the NSF and
NASA for support.  This work is based on observations obtained at the
W. M.  Keck Observatory, which is operated jointly by the University
of California and the California Institute of Technology.  The Munich
Image Data Analysis System is developed and maintained by the European
Southern Observatory.  Finally we thank Fustov Kirchoff and Robert
Bunsen for discovering cesium.

\newpage

%--------------------------BIBLIOGRAPHY---------------------------

\newpage 

\begin{figure}
\figurenum{1}
\plotone{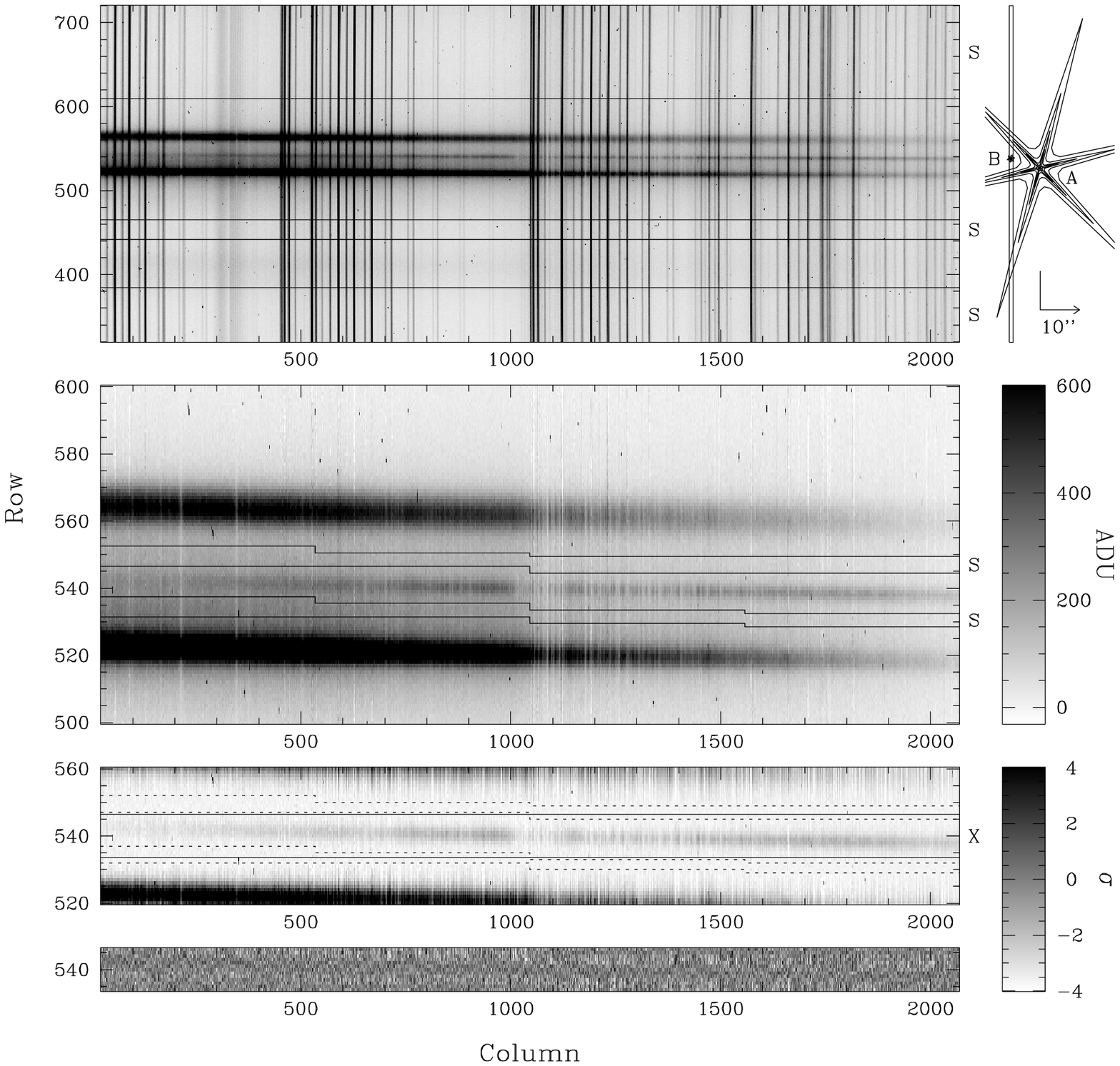}
\caption{\footnotesize Reduction of the optical spectra.  In the top
panel, one of the spectral images (a 900s integration) is reproduced,
after only bias subtraction and division by a normalized flat-field.
The whole spectral range is shown (blue to the left and red to the
right), but not the full length of the slit.  The grey scale is linear
between $-$30 and 600 ADU (the gain is 2.2\,e$^-$\,ADU$^{-1}$).  A
schematic view of the slit region of the spectrograph is also shown,
with Gliese 229B in the slit, and A nearby with six diffraction
spikes.  Spectra of three of the spikes can be seen in the image.  In
the middle panel, the data is shown after removing telluric sky
emission, as determined from fitting third-degree polynomials at every
column in the regions marked `S' to the right of the top panel.  The
contribution of scattered and diffracted light of Gliese 229A at the
position of B was determined by fitting second-degree polynomials at
every column in the two small regions marked `S' to the right of the
middle panel.  The corrected data is shown in the third panel.  The
absorption due to steam bands, near column 1000, is clearly seen.  The
spectrum was extracted from the region indicated by the two solid
lines, marked `X' at the right-hand side.  The fluxes weighted
according to a spatial profile (see text) are used to extract the
data.  The residual after extracting the data is shown in the bottom
panel in units of standard deviations, to show that the extraction is
accurate.  The scale in the Y direction is the same as that of the
third panel.  The grey scale is linear between $-4\sigma$ and
$4\sigma$, as indicated.}
\end{figure}

\begin{figure}
\figurenum{2}
\plotone{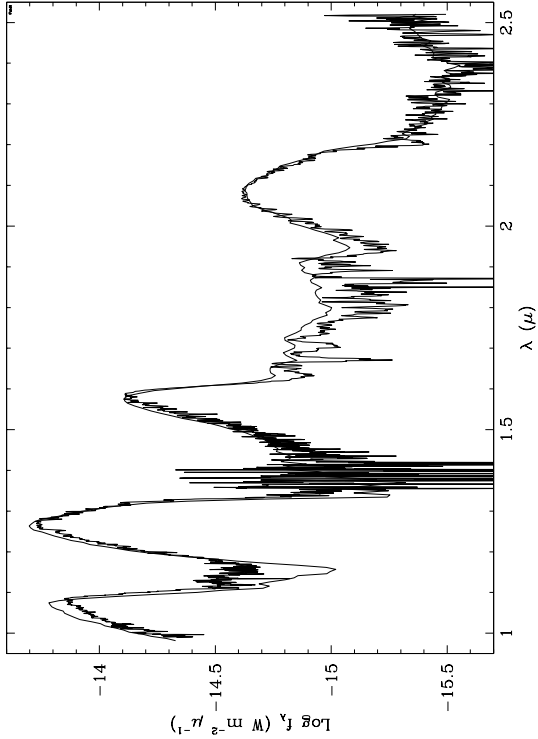}
\caption{\footnotesize The Keck NIRC spectrum of Gliese 229B (thick)
compared with the UKIRT CGS4 spectrum (thinner, higher resolution).
The agreement between the two spectra is good except in the 1.6 to 2
$\mu$m region and the 1 to 1.1 $\mu$m region.  The discrepancies are,
however, no more than 10\%\ of the signal, and are most likely due to
small differences in wavelength calibration and possibly to
atmospheric dispersion, which could put some of the flux at the blue
end of the UKIRT spectrum just outside the slit.  This would explain
the fact that the discrepancies are mostly on the blue end of the
spectrum.  Alternaitively, there may be minor residual contamination
from Gliese 229A in our spectrum.  The UKIRT spectrum is taken from
Geballe et al. (\cite{geb96}).}
\end{figure}

\begin{figure}
\figurenum{3}
\plotone{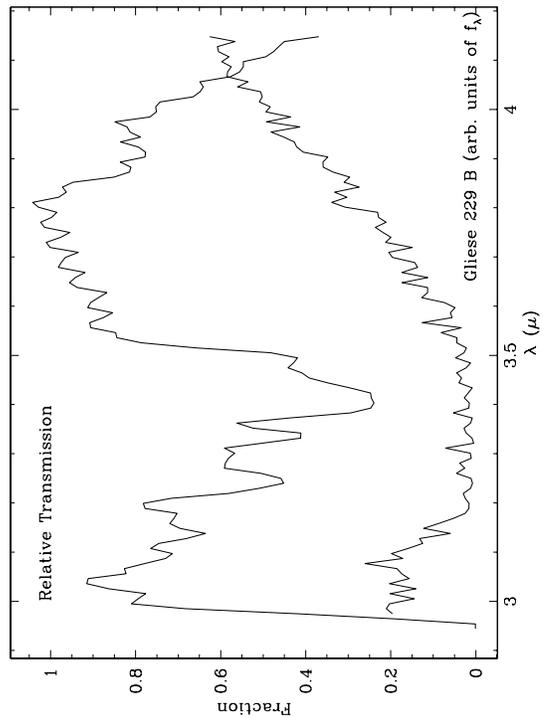}
\caption{\footnotesize The relative instrumental transmission of
NIRC through the L band (heavy line) and the f$_{\lambda}$ of Gliese
229B (lighter line).  The large variations in the instrumental
transmission are due to the resin that forms the rulings on the grism
in NIRC.  See \S \ref{sec:nirc} for a more detailed
discussion.}
\end{figure}

\begin{figure}
\figurenum{4}
\plotone{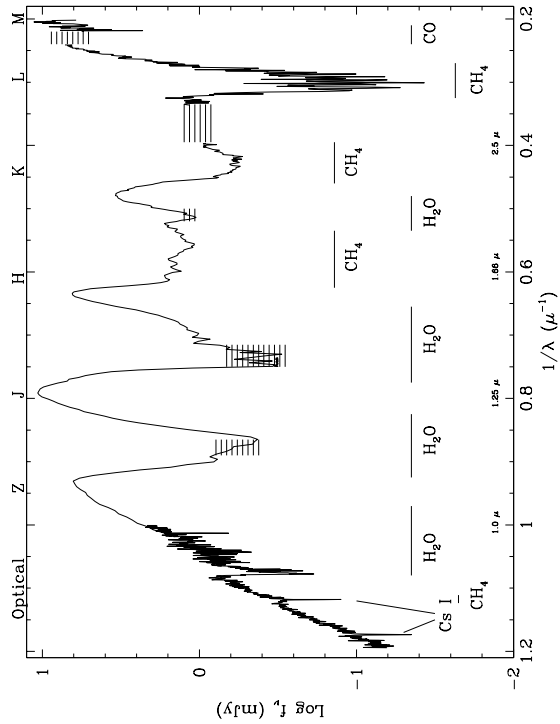}
\caption{\footnotesize The spectrum of Gliese 229B from 0.84
$\mu$m to 5 $\mu$m.  Major opacity sources are indicated.  Regions
with horizontal bars correspond to wavelengths where the atmosphere is
too opaque to permit collection of useful data from the ground.  Along
the top of the plot are indicated the filters corresponding to the
various wavelength bands.}
\end{figure}

\begin{figure}
\figurenum{5}
\plotone{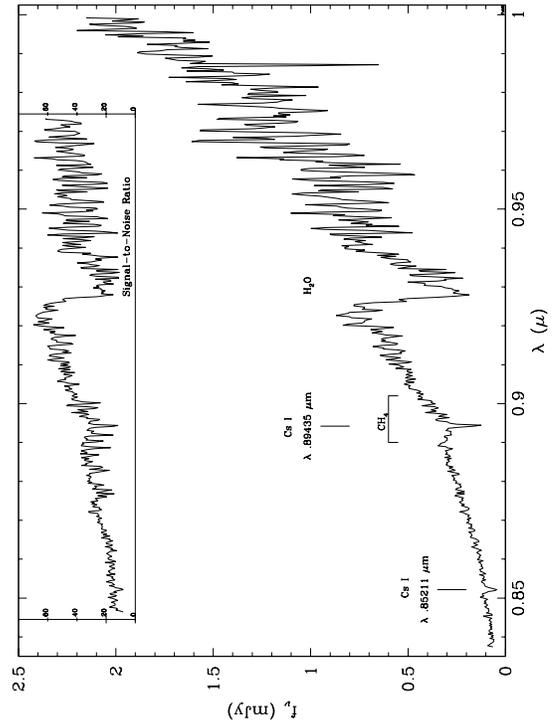}
\caption{\footnotesize The spectrum of Gliese 229B from 0.837
$\mu$m to 1.0 $\mu$m.  Important absorption features are labelled.
The upper panel is a graph of the signal to noise ratio through the
central region of the spectrum.  The redward end of the spectrum
degrades considerably in signal-to-noise ratio because of the
decreasing sensitivity in the CCD.  The seemingly noisy
long-wavelength end of this spectrum contains what are largely real
features.  The dips in signal to noise ratio near 0.88 $\mu$m are due to
telluric emission lines.}
\end{figure}

\begin{figure}
\figurenum{6}
\plotone{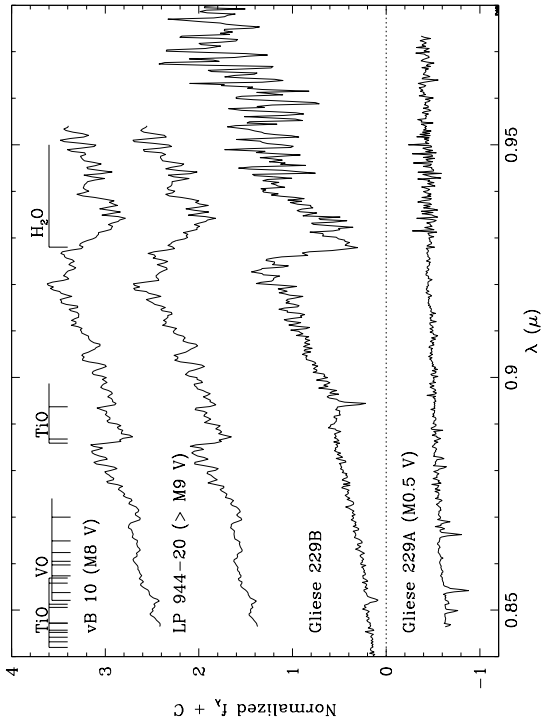}
\caption{\footnotesize The optical spectrum of Gliese 229B compared
with spectra of two of the least massive stars known, vB 10, an M8 V
star, and LP 944$-$20 whose spectral classification is $\gtrsim$ M9 V
according to Kirkpatrick et al. (\cite{kp97}).  These spectra are
presented here courtesy of J. D. Kirkpatrick and appeared earlier in
Kirkpatrick et al. (\cite{kp97}).  The values of C are 1.0 for LP
944$-$20 and 2.0 for vB 10.  Line locations, with the exception of Cs
and CH$_4$ are from Kirkpatrick et al. (\cite{kp91}).}
\end{figure}

\begin{figure}
\figurenum{7}
\plotone{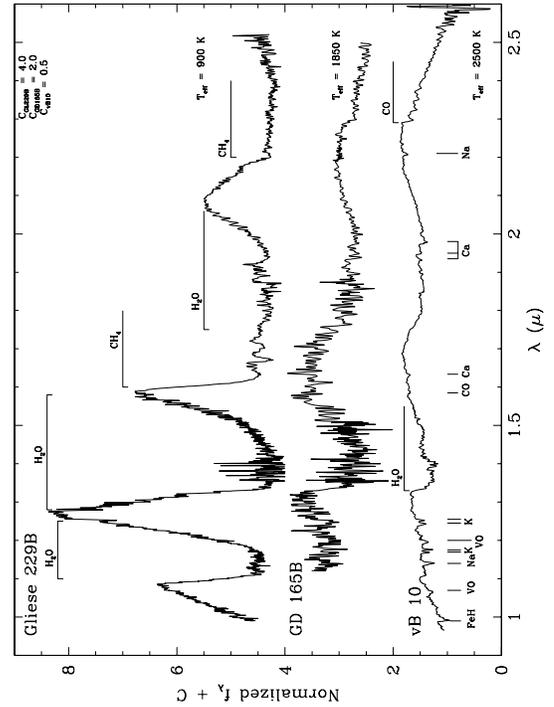}
\caption{\footnotesize The near IR spectrum of Gliese 229B compared
with spectra of vB 10 and GD 165B.  The IR spectrum is that of Geballe
et al. (\cite{geb96}) and is used because it matches the resolution of
the other two spectra which appear courtesy of H. R. A. Jones and were
first published in Jones et al. (\cite{jon94}).  The importance of
methane in the spectrum of Gliese 229B is best demonstrated in this
figure.  The distinct drop-off in flux density at 1.6 and 2.2 $\mu$m
are features not to be found in stellar spectra.  The water bands at
1.1 and 1.4 $\mu$m are also grossly magnified in strength in Gliese
229B.  The absorption line identifications for vB 10 are from Jones et
al. (\cite{jon94}).}
\end{figure}

\begin{figure}
\figurenum{8}
\plotone{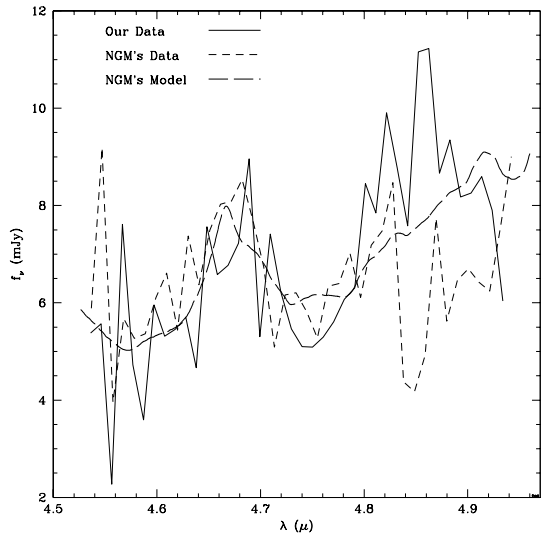}
\caption{\footnotesize Expanded view of the spectrum of Gliese 229B
between 4.5 and 5.0 $\mu$m.  The solid line represents the data
presented in this paper, while the short-dashed line comes from Noll
et al. (\cite{nol97}), as does the model spectrum represented by the
long-dashed line.  Typical uncertainties in our data points are about
\twothirds\ those of Noll et al. (\cite{nol97}), which are on the
order of 2 mJy.  Our data confirm the detection of carbon monoxide in
Gliese 229B's atmosphere.  The peak at 4.67 $\mu$m is due to a gap in
the middle of the 1-0 vibration-rotation absorption band of CO.  The
model uses a mole fraction of qCO = 200 ppm and a brown dwarf radius
of 80000 km.  The feature at 4.85 $\mu$m in the Noll et
al. (\cite{nol97}) data is due to incorrect removal of a telluric
line.  Their data and model appear courtesy of K. Noll, T. Geballe and
M. Marley.}
\end{figure}

\end{document}